\documentclass[12pt]{article}
\usepackage{acronym}
\usepackage{graphicx}
\usepackage{amssymb}
\usepackage{epstopdf}
\usepackage{amsmath}
\usepackage{bbm}
\usepackage{algorithmic}
\usepackage{subfigure}
\newcommand{\ket}[1]{|#1\rangle}

\newcommand{\braket}[2]{\langle#1|#2\rangle}

\renewcommand{\b}{\bar}
\renewcommand{\Pr}{P}

\newtheorem{theorem}{Theorem}[section]
\newtheorem{proposition}{Proposition}[section]
\newtheorem{corollary}{Corollary}[section]

\acrodef{IR}[IR]{Information Retrieval}
\acrodef{BP}[BP]{Bayes' Postulate}
\acrodef{BT}[BT]{Bayes' Theorem}
\acrodef{LTP}[LTP]{Law of Total Probability}
\acrodef{QM}[QM]{Quantum Mechanics}
\acrodef{LM}[LM]{Language Modelling}
\acrodef{QP}[QP]{Quantum Probability}
\acrodef{CP}[CP]{Classical Probability}
\acrodef{QS}[QS]{Quantum Space}
\acrodef{CS}[CS]{Classical Space}
\acrodef{MAP}[MAP]{Mean Average Precision}

\title{When Index Term Probability Violates the \acl{CP} Axioms \acl{QP} can be a Necessary
  Theory for \acl{IR}}
\author{Massimo Melucci \\ University of Padua \\ Italy}
\date{m.melucci@acm.org}

\begin{document}

\maketitle

\begin{abstract}
  Probabilistic models require the notion of event space for defining a probability
  measure. An event space has a probability measure which ensues the Kolmogorov
  axioms. However, the probabilities observed from distinct sources, such as that of relevance
  of documents, may not admit a single event space thus causing some issues. In this article,
  some results are introduced for ensuring whether the observed probabilities of relevance of
  documents admit a single event space. Moreover, an alternative framework of probability is
  introduced, thus challenging the use of classical probability for ranking documents. Some
  reflections on the convenience of extending the classical probabilistic retrieval toward a
  more general framework which encompasses the issues are made.
\end{abstract}

\section{Introduction}
\label{sec:introduction}

In \ac{IR}, probabilistic models are employed for estimating the probability
that a relevance or generation relationship exists between the information conveyed by a
document and a user's information need represented by a query or any other user's action, such
as browsing, document retention or click-through. These models requires an event space, which
consists of a set of events and a probability measure, thus two event spaces differ due to
either the events or the probability distribution.

Sets of events are employed for representing the occurrence of documents, terms, queries,
relevance, aboutness, and their inter-relationships; for example, the intersection of the set
which represents a document and the set of relevance represents the event that the document is
relevant. For every single event space, a probability distribution $\Pr$ exists such that, for
any pair of events, we can write the conditional probabilities as
\begin{equation}
  \label{eq:1}
  \Pr(A|B) = \frac{\Pr(A \cap B)}{\Pr(B)}
\end{equation}
The latter is known as \ac{BP}, which is different from \ac{BT} which states that, for any
pairs of events,
\begin{equation}
  \label{eq:3}
  \Pr(B)\Pr(A | B) = \Pr(B | A)\Pr(A)
\end{equation}
While \ac{BP} is a postulate, \ac{BT} permits to compute probability distributions through the
probability update mechanism provided by~\eqref{eq:3}. Note that \ac{BP} implies \ac{BT} but
the \textit{vice versa} does not hold. Finally, and mostly important, the events whose
probabilities used to compute the conditional probabililities through \ac{BP} belong to a
single probability space.

Some simple numerical examples in~\cite[pages 320, 321]{Robertson05} point out that there
might be something wrong or imprecise when defining the event space.  In particular, some
contradictions were noted when conditional probabilities are estimated from distinct event
spaces and are then combined together by using \ac{BP} since as the events whose probabilities
used to compute the conditional probabililities through \ac{BP} come from a single probability
space.  It was noticed that \ac{BP} cannot be applied without asking some questions and that
some measurements expressed in terms of events or random variables are simply ill-defined.

The contradictions found in~\cite{Robertson05} can be briefly explained as follows. Suppose
that we are given two conditional probabilities $\Pr(A | B), \Pr(A | C)$ calculated by using
\ac{BP}, and we are also told that $A,B,C$ do not necessarily come from a single event
space. In other words, these probabiilities are {\em estimated} under different
\emph{contexts} -- $\Pr(A | B)$ has been estimated under $B$ and $\Pr(A|C)$ has been estimated
under a {\em different} context $C$ where the conditioning events (i.e., $B$ and $C$ are the
contexts). Hence a single measure $\Pr$ may not exist such that~\eqref{eq:1}
holds~\cite{Accardi84}.

Another question is whether \ac{BT} may still hold when the probabilities are estimated under
different contexts. Suppose one is provided with an estimation of $\Pr_1(B|A),
\Pr_4(B|\b{A}), \Pr_0(A)$ from three different event spaces\footnote{The subscripts suggest
that the $\Pr$'s refers to different event spaces.}. By summing the
right-hand side of~\eqref{eq:3} over $A$, we have that 
\begin{equation}
  \label{eq:2}
  \Pr_2(B) = \Pr_1(B|A)\Pr_0(A) + \Pr_4(B|\b{A})\Pr_0(\b{A})
\end{equation}
whose sum over $B$ yields $1$, the latter being called \ac{LTP}. Hence, the following
probability distributions are provided:
\begin{itemize}
\item $\Pr_0(A)$ from event space $0$,
\item $\Pr_1(B|A)$ from event space $1$,
\item $\Pr_2(B)$ from event space $2$,
\item $\Pr_3(A|B)$ from event space $3$,
\item $\Pr_4(B|\b{A})$ from event space $4$.
\end{itemize}
However, one cannot choose the probability distributions at any degree of freedom.  Some
inequalities constrain the set of admittable probability distributions. Consider $\Pr_3(A|B) +
\Pr_3(\b{A}|B) = 1$.  Hence, for every $\Pr_0(A)$, we have that $\Pr_1(B|A) \leq \Pr_2(B)
\leq \Pr_4(B|\b{A})$ or $\Pr_4(B|\b{A}) \leq \Pr_2(B) \leq \Pr_1(B|A)$ or equivalently
\begin{equation}
  \label{eq:6}
  \left\|\frac{\Pr_2(B) - \Pr_1(B|A)}{\Pr_4(B|\b{A})  - \Pr_1(B|A)}\right\| \leq 1
\end{equation}
Inequality~\eqref{eq:6} is named statistical invariant in~\cite{Accardi84} and is necessary
and sufficient condition for the existence of a valid $\Pr(B)$. In particular, when the event
space is supposed to be Boolean and then event intersection such as $A \cap B$ can be
appropriately defined and observed, the violation of~\eqref{eq:6} tests the existence of a
single event space equipped with event intersection. If~\eqref{eq:6} was not admitted, one
cannot state that the observed $\Pr_2(B), \Pr_1(B|A), \Pr_4(B|\b{A}$ come from a single
event space, and \ac{LTP} is violated. 

It seems that a violation of~\eqref{eq:6} is detriment to \ac{IR} effectiveness because, one
may suppose, a ``wrong'' model of the world cannot provide nothing but ``wrong'' results. In
contrast to intuition, as it often happens, some experiments have shown that the violation can
lead to improvements of retrieval effectiveness~\cite{Melucci10}. These results are briefly
described in the following section.

\section{An Experiment Violating Some Probability Axioms}
\label{sec:an-exper-viol}

It was observed that the terms that correspond to $B,C$ and violates~\eqref{eq:6}, are those
that increase average precision more frequently and significantly than those do not. (Event
$A$ means relevance.) The experiments reported in~\cite{Melucci10} aimed at answering the
following question: ``If the term suggested by the system to the user to expand the original
query was so that the probability of occurrence violates the \ac{LTP}, is the retrieval
effectiveness measured on the new list of retrieved documents higher than that measured on the
original list of retrieved documents?''.

\begin{figure}[t]
  \centering
  \includegraphics[width=1.0\textwidth]{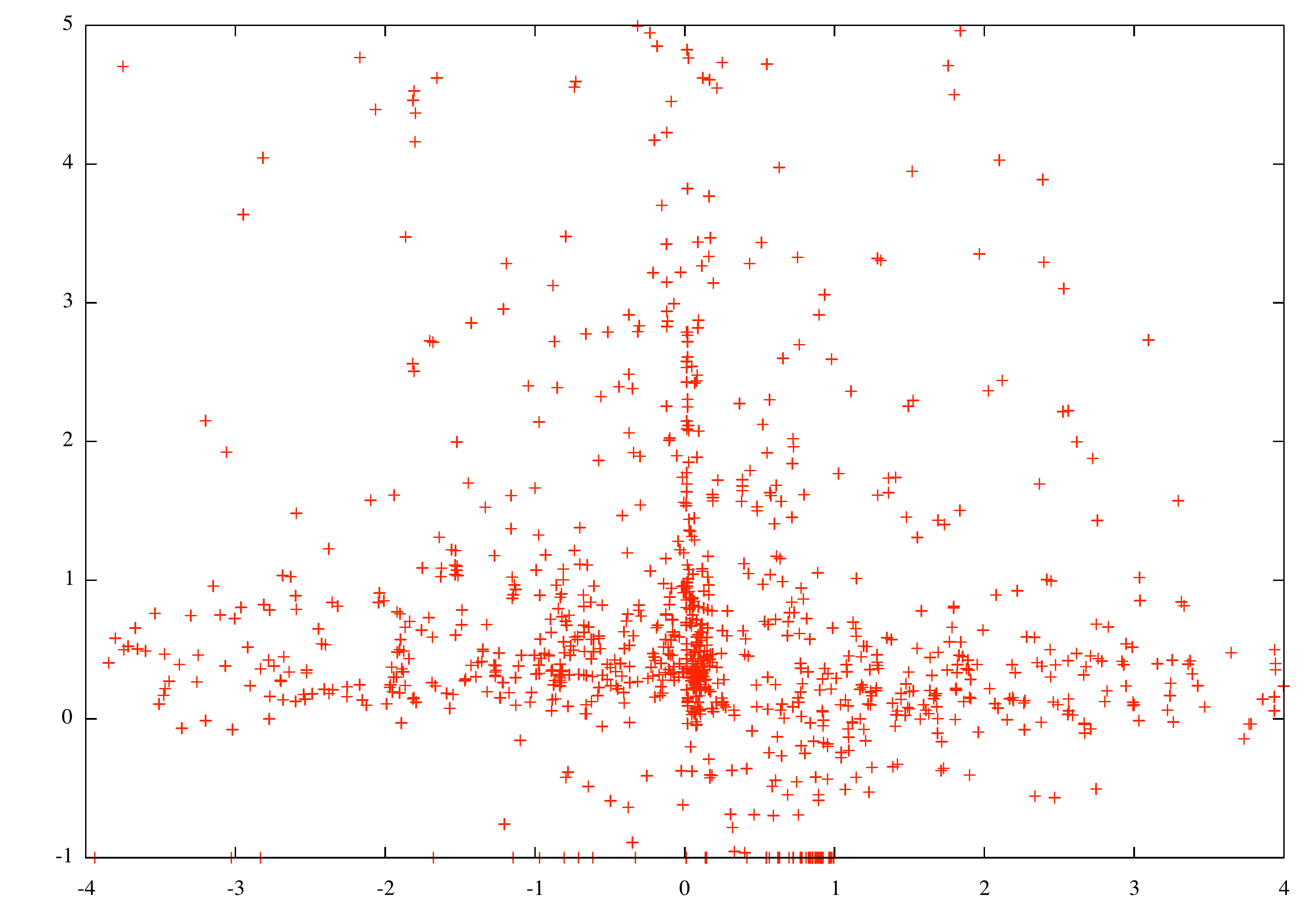}\label{fig:scatterplot-n20-k10-A}
  \caption{A scatterplot of \eqref{eq:6} and the increase of \ac{MAP}. Details in~\cite{Melucci10}}
  \label{fig:scatterplot-n20-k10-A}
\end{figure}

The experiments performed in~\cite{Melucci10} have suggested a relationship between the
variation in Mean Average Precision and the statistical invariant~\ref{eq:6} as depicted in
Figure~\ref{fig:scatterplot-n20-k10-A}. A point of this plot corresponds to a value of
fraction of \eqref{eq:6} and an increase of \ac{MAP}. The plot suggests that if the
statistical invariant cannot be admitted and the LTP is violated, that is, if the selected
term cannot be drawn from a sample whose probabilities have been estimated by aa single event
space built from a set of relevant and non-relevant documents, then the increase in retrieval
effectiveness may be observed and not only when the probability of occurrence of the selected
term admits the invariant. 

This experimental observation naturally lead to the question whether behind the violations of
seemingly necessary theoretical invariants, are some potential for even further improving
\ac{IR} effectiveness. Something like this hypothesis has been formulated, for example,
in~\cite{vanRijsbergen04} where the unification of logic, probability and vector space
geometric within a single non-classical framework shapted by \ac{QM} has been suggested as a
useful direction to this end. It follows that, other inequalities can be formulated that their
violation can reveal, from the one hand, the inconsistency between experimental observations
and the model, but, on the other hand, the directions toward a further development of
probabilistic models for \ac{IR}. The inequalities used throughout this article are
illustrated in Appendix~\ref{app:inequalities} and are based on~\cite{Accardi&82}. In spite of
the title, the contribution of~\cite{Accardi&82} is not only an explanation of the role played
by complex numbers in Quantum Mechanics, but it gives sufficient and necessary conditions for
the admissibility of the conditional probabilities by a \ac{CP} or \ac{QP}
space. The topic was further investigated in~\cite{Accardi97}.

\section{Inequalities of Probability and \ac{IR}}
\label{sec:inequalities-ir}

An \ac{IR} system is trained on how to rank documents on the basis of the past
interactions between the user and the system. These training data are recorded, for example,
in a log-file or observed during a session in which the user searches for documents which fill
his own information need.  A notable example of training is Relevance Feedback and its several
variations.  When a system is trained by feedback the training data can be described as a
table. Each tuple is an elementary event and one value is observed for each attribute and for
each elementary event; for example, a tuple refers to a relevant document and the other
attributes refer to the presence or absence of index terms\footnote{Tables is the standard way
  for representing training data as argued by I.H. Witten and E. Frank, \emph{Data Mining:
    Practical Machine Learning Tools and Techniques}, Morgan Kaufmann, 2005.}.

For the sake of clarity and for making the illustration close to an IR scenario, three
observables $A,B,C$ are considered in this section; one can imagine that $A$ refers to a set
of relevant documents, $B$ refers to the set of documents indexed by a term and $C$ refers to
the set of documents indexed by another term.  This simplification does not eliminate the
generality of the results because it will in the following be shown that some important
properties will be violated even in this simple scenario, and therefore will in general be
violated. At any rate, the most general case will be addressed in Section~\ref{sec:prp}.  The
set of tuples of the table which stores the training data can represent the event
space. Therefore, $A$ is both an attribute and the subset of elementary events with that
attribute; similarly, $A \cap B$ is another subset of elementary events.  
\begin{figure}[t]   
  \small
  \subfigure[][Elementary events]{  
    \begin{minipage}[t]{1.0\linewidth}
      \begin{equation*}
        \leavevmode
        \begin{array}[t]{cccccccccc}
          1	& 2	& 3	& 4	& 5	& 6	& 7	& 8	& 9	& 10 \\
          A	& \b{A}	& \b{A}	& \b{A}	& \b{A}	& A	& \b{A}	& A	& A	& A\\ 
          B	& B	& B	& B	& \b{B}	& B	& \b{B}	& \b{B}	& \b{B}	& \b{B}\\ 
          C	& C	& C	& C	& C	& \b{C}	& \b{C}	& \b{C}	& \b{C}	& \b{C}\\ 
        \end{array}
      \end{equation*}
    \end{minipage}
    \label{fig:log-file-0a}
  }  
  \subfigure[][Event space]{  
    \begin{minipage}[t]{1.0\linewidth}
      \begin{equation*}
        \leavevmode
        \begin{array}[t]{cccccccc}
          A	& A	& A	& A	& \b{A}	& \b{A}	& \b{A}	& \b{A} \\ 
          B	& B	& \b{B}	& \b{B}	& B	& B	& \b{B}	& \b{B} \\
          C	& \b{C}	& C	& \b{C}	& C	& \b{C}	& C	& \b{C} \\
          1	& 1	& 0	& 3	& 3	& 0	& 1	& 1	\\
        \end{array}
      \end{equation*}
    \end{minipage}
    \label{fig:log-file-0b}
  }  
  \caption{Event space for three observables.}
  \label{fig:log-file-0}
\end{figure}

An example is illustrated in Figure~\ref{fig:log-file-0}. The set of tuples of
Figure~\ref{fig:log-file-0a} for which both terms occur, i.e., $\{1,2,3,4\}$, is an event
labeled $A,B,C$ and placed in the first column of Figure~\ref{fig:log-file-0b}; the last
number of a column of Figure~\ref{fig:log-file-0b} is the cardinality of this event and it is
used for computing the probability that both terms occur~--~the tuples that do not occur
(e.g. $A, \b{B}, C$) have null probability.  When moving to the conditional probabilities,
$p=\Pr(B | A)$ is the probability that a term has been observed in a relevant document,
$r=\Pr(C | A)$ is the probability that the other term has been observed in a relevant
document, and $q=\Pr(C | B)$ is the probability that a term has been observed when the other
term has also been observed. Suppose also that the frequencies of Figure~\ref{fig:log-file-0b}
are used for estimating the probabilities by using \ac{BP}. The example of
Figure~\ref{fig:log-file-0} obviously is a single event space and indeed
Inequality~\ref{eq:accardi-1} holds. 

Inequalities might be violated if estimation is based on different event spaces.  In IR, the
violation of Inequality~\ref{eq:accardi-1} may happen, for example, when
\begin{enumerate}
\item $p,q,r$ are estimated using a mixture
\item $p,q,r$ are estimated from distinct collections, or
\item the value of an observable is missing in the data for a few tuples. 
\end{enumerate}
When the probabilities are estimated by a mixture of, say, frequencies and additional
knowledge, different sources are combined for estimating the probabilities $p,q,r$. This may
happen, for example, in the \ac{LM} approach when other sources of evidence, such as
additional log-files or collection term frequency distributions, are exploited for adding some
parameters and for smoothing probabilities.  A linear combination is a well-known method for
smoothing probabilities thus obtaining $p,q,r$ as follows:
\begin{eqnarray}
  p = \alpha\frac{1}{2}+(1-\alpha)\frac{3}{4} \qquad 
  q = \beta\frac{1}{2}+(1-\beta)\frac{1}{4} \qquad 
  r = \gamma\frac{1}{2}+(1-\gamma)\frac{9}{15}\ .
  \label{eq:example-3}
\end{eqnarray}
Suppose that 
%\begin{eqnarray*}
%  a_p = 3 \qquad b_p = 1 \qquad a_q = 1 \qquad b_q = 3 \qquad a_r = \frac{9}{4}
%  \qquad b_r = \frac{3}{2}\ .
%\end{eqnarray*}
\begin{eqnarray*}
  \alpha = \frac{1}{9} \qquad \beta = \frac{1}{9} \qquad \gamma =
  \frac{2}{17}
\end{eqnarray*}
By using Inequality~\ref{eq:accardi-1}, one can check that $p,q,r$ do not admit a single event space and \ac{BT} cannot be applied for computing $\Pr(A | B)$ and $\Pr(A | C)$.

A similar case happens when $p,q,r$ are estimated from distinct collections; in this case, the
experimental conditions yielding the probabilities are different and cannot be compared
although the relevance assessments were given in the best possible way. This may happen, for
example, when a document is stored in a collection, another document is stored in another
collection and the query is routed to both these collections for retrieving and ranking the
two documents in a single list. One may also think about routing the query to a collection
drawn at random, or about merging the results received from a collection with those received
from another after weighing the probabilities of relevance used for ranking the single lists.

\begin{figure}[t]   
\small
  \begin{equation*}
    \leavevmode
    \begin{array}[t]{cccccccccc||cccccccccc}
      \multicolumn{10}{c}{S_1} & \multicolumn{10}{c}{S_2}\\
      1 & 2 & 3 & 4 & 5 & 6 & 7 & 8 & 9 & 10 &  1 & 2 & 3 & 4 & 5 & 6 & 7 & 8 & 9 & 10\\
      A & A & A & A & A & \b A & \b A & \b A & \b A & \b A & A & A & \b A & A & A & \b A & \b A & \b A & \b A & A\\ 
      B & B & \b B & \b B & \b B &  B & B & B & \b B & \b B &  B & B & \b B & \b B & \b B &  B & B & B & \b B & \b B\\ 
     C & \b C & C & \b{C} & \b{C} & \b C & \b C & \b C & C & \b C  & C & \b C & C & \b{C} & \b{C} & \b C & \b C & \b C & C & \b C\\ 
    \end{array}
  \end{equation*}
  \caption{Ten elementary events for which two observables were measured for each urn.}
  \label{fig:log-file-2}
\end{figure}
Suppose, for example, that ten documents are stored in collection $S_1$ and other ten documents are stored in collection $S_2$~---~Figure~\ref{fig:log-file-2} reports an example.
Let $p_i, q_i, r_i$ be the three conditional probabilities observed from $S_i$ for $i=1,2$. 
One can easily check that
\begin{equation*}
p_1 = \Pr(B | A) = \frac{2}{5} \qquad q_1 = \Pr(B | C) = \frac{1}{5} \qquad r_1 = \Pr(C | A) = \frac{2}{5} 
\end{equation*}
and that
\begin{equation*}
p_2 = \Pr(B | A) = \frac{2}{5} \qquad q_2 = \Pr(B | C) = \frac{1}{5} \qquad r_2 = \Pr(C | A) = \frac{1}{5} 
\end{equation*}
The three conditional probabilities estimated from $S_i$ do admit a single event space for each $i$~---~indeed, one single measure $\mu(X)$ can be defined as the frequency of the elementary events in $X$ for each subset $X$ of $S_i$. 

Suppose that a query has to be submitted to a broker which has to decide the $S_i$ to which the query has to be routed. The broker may either pick a collection at random with probability $\alpha$ and then route the query, or to route the query to both of them and then weigh the probabilities with weight $\alpha$. In both cases, the conditional probabilities observed from the two collections may be
\begin{equation*}
p = \alpha p_1 + (1-\alpha)p_2 \qquad q = \alpha q_1 + (1-\alpha)q_2 \qquad r = \alpha r_1 + (1-\alpha)r_2\ .
\end{equation*}
In this way, $p$ is an estimation of the probability that $B$ is observed in a relevant document stored in the $S_i$'s where $\alpha$ is the prior probability that $S_1$ is selected. Similarly, $q,r$ are estimated as a mixture of the conditional probabilities estimated from the single collections.

By using Inequality~\ref{eq:accardi-1}, one can check that when $\alpha = \frac{1}{2}$, the conditional probabilities 
\begin{equation*}
p = \frac{4}{10} \qquad q = \frac{2}{10} \qquad r=\frac{3}{10}
\end{equation*}
do not admit a single event space. As a consequence, one cannot compute, say, $\Pr(A | B)$ from the probabilities of the broker by using \ac{BT} even though \ac{BT} could be used for the probabilities estimated from single $S_i$. The reason was that the probabilities $p_i,q_i,r_i$ come out from distinct spaces which describe two different experimental conditions. However, two distinct $S_i$'s may still permit the probability of relevance to be computed by mixing the $p_i,q_i,r_i$'s if appropriate values of $\alpha$ are fixed.

\begin{figure}[t]
\small
  \begin{equation*}
    \leavevmode
    \begin{array}[t]{cccccccccccc}
      A & \b{A} & \b{A} & \b{A} & \b{A} & A & \b{A} & A & A & A&?&?\\ 
      B & B & B & B & \b{B} & B & \b{B} & \b{B} & \b{B} & \b{B}&B&\b{B}\\ 
      C & C & C & C & C & \b{C} & \b{C} & \b{C} & \b{C} & \b{C}&C&\b{C}\\ 
    \end{array}
  \end{equation*}
  \caption{Twelve elementary events are another set of examples of the
  relevance (\{$A,\b{A}\}$), the presence of a term
  (\{$B,\b{B}\}$) and the presence of another term (\{$C,\b{C}\}$)
  observed in a log-file. Two elementary events include missing or
  known values for relevance.}
  \label{fig:log-file-1}
\end{figure}
Another situation when the conditional probabilities do not admit a single event space occurs in the event of unknown or missing values, as exemplified in Figure~\ref{fig:log-file-1}. 
Depending on how the conditional probabilities are estimated, a single event space holds or does not.  An estimation may only involve ten elementary events for which either $A$ or $\b A$ is known so that
\begin{equation*}
  p = \frac{\mu(B \cap A)}{\mu(B)} = \frac{2}{5} \qquad
  q = \frac{\mu(B \cap C)}{\mu(C)} = \frac{4}{5} \qquad
  r = \frac{\mu(C \cap A)}{\mu(C)} = \frac{1}{5}
\end{equation*}
Another estimation may also involve the tuples for which neither $A$ or $\b A$ is known so that
\begin{equation*}
  p = \frac{2/10}{6/12} = \frac{2}{5} \qquad
  q = \frac{5/12}{6/12} = \frac{5}{6} \qquad
  r = \frac{1/12}{6/12} = \frac{1}{6}
\end{equation*}
When the latter estimation is used, Inequality~\ref{eq:accardi-1} is violated and therefore the single event space cannot hold. This outcome is little surprising when using the table of Figure~\ref{fig:log-file-1} because the universe of the elementary events has two tuples for which a value is missing or unknown. A shrewd experimenter will avoid such a situation, yet it should be noted that a great deal of attention should be paid when conditional probabilities are provided by some ``black-box'' device or when the dataset includes missing values.

In the rest of this section, an example more extensive than the three-observable toy example is reported.  The example aims at illustrating how the mathematical concepts described  above can arise when designing an information retrieval experiment. In particular, it is shown how the estimation of a prior probability is a crucial step. The use of a small collection, such as the CACM of this example, is not detrimental to the generality of the results because even a small test collection or an experiment setting may contain a counter-example which invalidates the hypothesis of single event. The use of a larger collection would have provided more counter-examples.

The test collection was indexed so as to relate each one-keyword term to the documents in which it occurs. Before computing term frequencies, the stopwords provided with the test collection were removed from the documents. No stemming was computed. After indexing, the probability of relevance was first computed for each query as the relative frequency between the number of relevant documents and the total number of documents. Second, the probability of observing a term was computed as the relative frequency between the number of documents indexed by the term and the total number of documents.  Third, the terms whose probability of occurrence was equal to the probability of relevance were selected, for each query.  Finally, the probability of co-occurrence of every pair of two selected terms was computed, for each query.  For example, the term \texttt{infinity} occurs in four documents and co-occurs with \texttt{translates} in one document; both terms have the same probability of occurrence as the probability of relevance to queries \texttt{23} or \texttt{30} which have four
relevant documents.

In terms of probability, the universe of elementary events was the collection of documents. For each document, the property ``the document was indexed by a term'' and the property ``the document is relevant to a given query'' were observed; specifically, the latter property or observable was $A$, whereas $B, C$ refer to two terms observed in a document. Therefore, the probability of relevance was $\Pr(A)$ and the probability of observing a term was $\Pr(X)$ where $X = B$ or $X = C$. From these probabilities, the conditional probabilities were computed. Specifically, the probability that a relevant document was indexed by $C$, i.e. $\Pr(C | A)$ was computed as the relative frequency of relevant documents indexed by $C$. In the same way, the conditional probability of co-occurrence, i.e. $\Pr(B | C)$ was computed as the relative frequency of documents indexed by $C$ were indexed also by $B$. For example, only one relevant document (\#\texttt{2786}) was indexed by \texttt{infinity} and therefore $\Pr(C | A) = \frac{1}{4}$.

A sample of the results is reported in Table~\ref{tab:sample}.  The first row of the table
includes a simple example which admits a single event space since, whenever a term is observed
in a document, it is certain that the other term or relevance is observed too. Given a query,
two terms whose probability of occurrence equals the probability of relevance are associated
with the conditional probability of co-occurrence and the conditional probability of
relevance. \ac{CS} stands for ``it admits a \acl{CS}'', Re\ac{QS} stands for ``it admits a
real \ac{QS}'' and Co\ac{QS} stands for ``it admits a complex \ac{QS}''.  When a term admits a
single event space, it admits a complex \ac{QS} too, yet a real one may not be admitted as
shown by the second row\footnote{\ac{QP} spaces are introduced in Appendix~\ref{app:hilbert}
  and are here mentioned to stress the fact that three observed conditional probabilities may
  admit a variety of theories of probability and not only the single event.}.  The
probabilities of the third row admit a real \ac{QS}, and then a complex one, while the single
event space is not admitted. The last one is an instance that a \ac{QS} can not be sufficient
and that another space should be defined.

\begin{table}[t]
  \begin{center}
    \footnotesize
    \leavevmode
    \begin{tabular}[t]{|c|l|l|c|c|c|c|c|c|}
      \hline
      Query & \multicolumn{2}{c|}{Terms}  & $\Pr(B|C)$    & $\Pr(B|A)$    & $\Pr(C|A)$    & \ac{CS}     & Re\ac{QS}     & Co\ac{QS} \\
      \hline
      33    &nonnormal     &attainable    &1            &1             &1             &Yes   &Yes   &Yes\\[4pt]
      30    & infinity      & typesetting   &$\frac{1}{4}$          &$\frac{1}{4}$          &$\frac{1}{2}$            &Yes    &No     &Yes\\[4pt]
      30    &translates     &infinity       &$\frac{1}{4}$          &$\frac{1}{4}$           &$\frac{1}{4}$           &No     &Yes    &Yes\\[4pt]
      37    &registers      &compatible     &$\frac{1}{12}$           &$\frac{1}{12}$           &$\frac{1}{12}$           &No     &No     &No\\[4pt]
      \hline
    \end{tabular}
  \end{center}
  \caption{A sample of the conditional probabilities computed for the   terms selected from the CACM test collections is reported in this  table. }
  \label{tab:sample}
\end{table} 

Let us concentrate on the third row.  If one compiled a table which enumerates all the elementary events~---~one event for each document~---~then each row of that table would include a binary value which indicates the presence/absence of a term, another binary value which indicates the presence/absence of the other term, the document identifier and a binary value which indicates whether the document is relevant or not. If such a row is added for each document, an $N$-row table is obtained, where $N$ is the total number of documents. One would easily check that a probability measure can be defined and therefore would find that a single event space can be defined. So, why does Table~\ref{tab:sample} say that K is not admitted?  The reason is due to the fact that Accardi's inequalities described in Appendix~\ref{app:inequalities} can be applied only if $\Pr(A) = \Pr(\b A) = \frac{1}{2}$ (and similarly states the same for $B$ and $C$), the latter being a hypothesis valid only if $N=8$.  This means that those inequalities must be checked only if $\Pr(A) = \Pr(\b A) = \frac{1}{2}$. 

%The lesson given is that the hypothesis made when estimating probabilities and applying \ac{BT} should carefully be checked. In particular, the probabilities of relevance are often estimated {\it a priori} and are considered as a neutral parameter in some cases. Indeed, it is often neglected because it is independent of the document and therefore does not affect ranking which depends on the likelihood ratio only. However, the computation of the likelihood ratio implies \ac{BT} which may not be admitted.

%In Section~\ref{sec:hilbertian-ir}, an alternative probability is proposed to overcome the lack of comparability between probabilities which do not admit a \ac{CP} space.

%It is in the following illustrated another situation when the
%incompatibility of the observables occur in IR. In particular, it is
%argued that the violation of the inequalities occurs more frequently
%than expected because of the design of the experiments and that this
%incompatibiblity is at the roots of the IR process thus making the use
%of \ac{CP}~---~and the probabilistic frameworks
%based on that probability~---~inappropriate for retrieval modeling.

\section{Inequalities of Probability and the Probability~Ranking~Principle}
\label{sec:prp}

What is the impact of the results illustrated in the previous sections on the PRP? Suppose that two documents have to be ranked. One document is represented by $B$, the other by $C$. When the observed conditional probabilities do not admit a single event space, the probability of relevance of $B$ cannot be confronted to the probability of relevance of $C$. When these two probabilities cannot be confronted, the ranking is questionable.  An explanation is that the probabilities are hinged on different measures, each defined by a distinct single event space which is an abstraction of a distinct experimental condition.  In order to make the ranking sound, a single measure is needed, which is provided by a single probability space such as a \ac{QS}. However, a single probability space is not sufficient for applying \ac{BT}~---~for example, \ac{BP} cannot be invoked if a \ac{QS} is used.  These issues are addressed in the rest of the article; in particular,  the general case of an arbitrary number of documents to be ranked and then the impact on the PRP are investigated.  

Let us now consider the general case. Suppose $m = n-1$ documents are to be ranked by the probability of relevance~---~relevance is then the $n$-th observable. The events corresponding to the documents are $D_1, \ldots, D_{m}$, whereas the event corresponding to relevance is $A$, so the estimated probabilities are $\Pr(D_i | A)$ for all $i=1,\ldots,m$. The correlation vector of probabilities $\textbf{p}$ used in Pitowsky's Theorem~\ref{prop:pitowsky-0} can be built as follows:
\begin{eqnarray*}
  p_i     &=&\Pr(D_i) \\
  p_{i,n} &=&\Pr(D_i | A)\Pr(A) \qquad i=1,\ldots m \\
  p_n     &=&\Pr(A)
\end{eqnarray*}

%Because the $\textbf{p}(\epsilon)$'s are
%fixed and cannot be changed, and the $\lambda$'s are the unknowns
%which cannot be changed, too, the only way to have a solution of the
%system is to change $\textbf{p}$. That is, the only way to have a
%\ac{CP} space is to have another vector of the probabilities
%observed from the devices so that the system can be solved. In order
%to change $\textbf{p}$, it is necessary to replace either a $p_i$ or a
%$p_{i,j}$. Let $\textbf{p}'$ be the new vector or probabilities.  As
%$\textbf{p}'$ makes the system solvable, a measure $\mu$ exists such
%that
%\begin{eqnarray*}
%  p'_i     &=&\mu(D_i) \\
%  p'_{i,n} &=&\mu(A \cap D_i) \qquad i=1,\ldots m \\
%  p'_n     &=&\mu(A)
%\end{eqnarray*}
%Suppose, for example, that $p'_1$ was the only new probability. 

Now, suppose that $\textbf{p}$ does not admit a single event space, that is, there is no set of values of $\lambda_1, \ldots, \lambda_{2^n}$ such that $\textbf{p}$ can be expressed as a linear combination of fixed correlation vectors built from the $2^n$ binary strings (see Appendix~\ref{app:pitowsky}). Suppose, then, a subset of the $n$ probabilities (or events) is selected and that the $n-1$ probabilities $p(1), \ldots, p(m-1),p(n)$ are considered. As a consequence, the bivariate probabilities $p(i,j)$ are computed from the corresponding events. In total, a $(n-1)n/2$-dimensional correlation vector $\textbf{p}'$ is observed. If this correlation vector admits a single event space, then two single event spaces will be found since the probability left apart admits one distinct space. If $\textbf{p}'$ does not admit it, then the other similar correlation vectors are built by leaving one of $p(1), \ldots, p(m-1)$ apart at a time.  If no $(n-1)n/2$-dimensional correlation vector admits a single event space, then the process is repeated by leaving two events apart until a single event space is admitted for $n-2$ events. The probability of relevance $p(n)=\Pr(A)$ can be left last. It is known that the process will certainly end finding a single event space because the single $p(i)$ does always admit it. This means that eventually every document or relevance may be represented as an observable of a space being distinct from all the spaces which represent the other documents or relevance. %The process is formalized in Figure~\ref{fig:proof} which proves the following
\begin{corollary}
  Let $\textbf{p}$ be a correlation vector for (not necessarily disjoint) $n$ events which
  does not admit a single event space. Then, there exist at least two subsets of events whose
  probabilities admit a single event space.
  \label{cor:proof}
\end{corollary}

% \begin{figure}[t]
%   \begin{minipage}[t]{0.98\columnwidth}
%     \begin{algorithmic}
      
%     \end{algorithmic}
%     \begin{tabbing}
%   \=\hspace{5mm}\=\hspace{5mm}\=\hspace{5mm}\=\hspace{5mm}\=\hspace{5mm}\kill
%       \> Let $A = \{A_1,\ldots,A_n\}$ be $n$ events (not necessarily disjoints)\\
%       \> Let $\textbf{p}=(p(1), \ldots, p(n),p(1,2),\ldots,p(n-1,n))$
%       be the correlation vector for $\textbf{A}$\\
%       \> Invoke Check($\textbf{p}, A$)\\
%       \\
%       \> Check($\textbf{q}, J$)\\
%       \> \textbf{begin}\\
%       \> \> \textbf{if} $|J|=1$ \textbf{then}\\
%       \> \> \> $\textbf{q}$ admits a \ac{CP} space\\
%       \> \> \textbf{else}\\
%       \> \> \> \textbf{if} $\textbf{q}$ is in a polytope\\
%       \> \> \> \> $\textbf{q}$ admits a \ac{CP} space\\
%       \> \> \> \textbf{else}\\
%       \> \> \> \> \textbf{forall} $A_j \in J$\\
%       \> \> \> \> \> Let $\textbf{p}$ be the correlation vector for $J \setminus A_j$\\
%       \> \> \> \> \> Invoke Check($\textbf{p}, J \setminus A_j $)\\
%       \> \> \> \> \textbf{end for}\\
%       \> \> \> \textbf{endif}\\
%       \> \> \textbf{endif}\\
%       \> \textbf{end}\\
%     \end{tabbing}
%   \end{minipage}
%   \caption{The proof of Corollary~\ref{cor:proof}.}
%   \label{fig:proof}
% \end{figure}

\section{Conclusions and Future Developments}

The main conclusion which can be drawn from Corollary~\ref{cor:proof} is that, whenever one has to rank $m$ documents by probability of relevance, a great deal of attention should be paid as to whether these probabilities admit a single event space or not. As mentioned above, the ranking of documents whose probability of relevance is computed from distinct spaces should be treated with due caution because the presence of distinct spaces would signal the use of different experimental conditions in which the probabilities of relevance were computed. Even if one decided to compare these probabilities despite the presence of distinct spaces, some properties of single event spaces, such as \ac{BT} or distributivity, cannot be used since they are grounded on a single space. 

It is our opinion that a new canvas is needed for probabilistic retrieval which overcomes the problems arising when some observed conditional probabilities cannot admit a single event space which is at the basis of the classical probabilistic retrieval models. Such a canvas should be based on \ac{QP} for some reasons explained in the rest of this section.

In IR, it was conjectured that ``if IR models are to be developed [...], without further empirical evidence to the contrary it has to be assumed that subspace logic will be non-classical.''~\cite{vanRijsbergen04}. The conjecture is the same as that made in Quantum Mechanics where Hilbert's spaces are taken as the theoretical framework for explaining how the phenomena studied by Physics happen in Nature at the particle-level. Whenever two observables $A,B$ are measured on a particle, the event $A \cap B$, that is, the event that both $A$ and $B$ occurs, often does not make sense, that is, the design of an experiment which can determine $A$ and $B$ cannot in principle be implemented because the measurement of $A$ interpheres with the measurement of  $B$ and thus nothing can be said about their co-occurrence.

If non-classical logic has to be assumed in IR too, the intersection of events cannot be assumed and as a consequence the events observed in IR cannot derive from the co-occurrence of events and then modeled as the intersection of sets. For example, one cannot observe both relevance and document as an event like $A \cap B$. 

Nonetheless, the phenomena usually dealt with in Physics at the particle-level do not at first sight occur in IR and the proposal of using Hilbert's spaces does not imply that IR systems exhibit a quantum behaviour; it rather means that the mathematical framework is general enough for describing documents, queries and the retrieval of relevant information in a comprehensive way. 

For measuring the uncertainty of the observation when non-classical logic has been supposed, \ac{QP} has been suggested. A probabilistic retrieval function based on \ac{QP} was illustrated in~\cite{Melucci08a} where a new model for information retrieval was proposed for capturing the contextual properties being hidden in an object managed by an IR system. According to that proposal, the contextual properties are modeled as bases of a complex vector space and each value, called contextual factor, taken by these properties is modeled as one of the basis vectors. The probability that a contextual factor occurs in an object was modeled as the square of the inner product between the vector which represents the factor and the vector which represents the object, and was termed as probability of context. However, some questions were left unanswered.

An unanswered question in~\cite{Melucci08a} was why the abstract vector spaces would be a better framework than other mathematical theories.  A possible answer was provided in~\cite{vanRijsbergen04}: Hilbert's spaces encompass different models for information retrieval, such as the probabilistic model and the VSM. That answer stemmed from the conjecture  that the use of non-classical logic is reasonable unless there is some evidence to the contrary. That notwithstanding, it was our opinion that it is still unclear why such a conjecture is necessary, and as a consequence, why \ac{QP} is necessary in IR and a further explanation was necessary.

The results of the previous sections may provide an explanation of why \ac{QP} is necessary in IR, that is, some conditional probabilities do not admit a single event space and therefore Bayes' results cannot be applied. Although \ac{QP} is necessary, it might unfortunately not be sufficient.

% \bibliographystyle{plain}
% \bibliography{general}

\appendix

\section{\acl{CP}}
\label{app:kolmogorov}

According to the \ac{CP}, the events observed during an experiment\footnote{Experiment is meant in the broad sense of a procedure for gathering a set of data, or the set of data itself, in the context of testing a hypothesis or studying a phenomenon.} or in the real world are modeled as \textit{sets}.  A set is nothing other than an abstraction of an event, that is, every event corresponds to a subset of a larger, perhaps infinite, universe of elementary events.  In this framework, the theory of sets together with its operations are an abstraction of the different ways the events can occur in the real world. In particular, the intersection of two sets models the conjunction, that is, the co-occurrence of two events, while the complement of a set models the negation of an event. 

When probability is entered, the sets used for modeling the events of a real world or an experiment are subjected to a measure, that is, a real function of the sets.  The measure is then used for computing the probability of the events modeled by the sets. In this way, the uncertainty of the occurrence of an event can effectively be measured. The event space together with the probability give rise to a probability space and in particular single event space. A notable example of the measure of a set (event) is the frequency of elementary events included by the the set (event), while its probability is the relative frequency.

Let us consider a mathematical formulation of the probability space, which will make things easier to illustrate in the remainder of this article. According to \ac{CP}, the events $A, B, \ldots$ are subsets of the event space $\Omega$ and therefore $A \cap B$, $A \cup B$ and $\b{A} = \Omega \setminus A$ are subsets of the event space, too. Moreover, suppose $\mu: 2^{\Omega} \rightarrow [0,1]$ is a measure of these subsets such that
\begin{equation*}
  \mu(\emptyset) = 0 \qquad 0 \leq \mu(A) \leq 1 \qquad \mu(\Omega) =
  1\ .
\end{equation*}
Note that $\mu$ is not necessarily a relative frequency. Moreover,
\begin{equation*}
  \Pr(A) = \mu(A) \qquad \Pr(A \cup B) = \Pr(A) + \Pr(B) \qquad \mbox{if} \qquad A \cap B = \emptyset 
\end{equation*}
According to \ac{BP}, which requires the single event space, the conditional probability of $A$ given $B$ is defined as:
\begin{eqnarray}
  \Pr(A  |  B) = \frac{\mu(A \cap B)}{\mu(B)}\ .
  \label{eq:k-prob}
\end{eqnarray}
Bayes' Theorem (\ac{BT}) states that
\begin{equation*}
  \Pr(A  |  B) = \frac{\Pr(B  |  A) \Pr(A)}{\Pr(B)}
\end{equation*}
The use of \ac{BT} entails \ac{BP} and then a single event space, and viceversa. In IR, the probabilities are often estimated through \ac{BT} which combines the probability that, for example, a query is generated by the document and that the query is generated ``a priori'' by the collection or another resource. 
%A special case of conditional probability is the following:
%\begin{eqnarray*}
%  \Pr(A  |  \Omega] = \frac{\mu(A \cap \Omega)}{\mu(\Omega)} = \mu(A)\ .
%\end{eqnarray*}
%This equality seems trivial, but it also highlights that the
%probability of an event, say, $A$, depends on the event space and on
%the measure. Indeed, the event space may include elementary events
%whose definition much derives from what should be modeled. For
%example, when the universe is discrete, the measure can be applied for
%every elementary event. In this way, the probability of an event $A$
%is the sum of the probabilities, i.e., the measures, over the
%elementary events of $A$. Thus, the probability of $A$ can change
%depending on the probability mass allocated to every elementary event.

%In IR, the role played by the event space in defining probabilistic
%retrieval models was illustrated in~\cite{Robertson05}.

\section{\acl{QP}}
\label{app:hilbert}

When Hilbert's spaces are considered, the events observed in an experiment or in the real world are modeled as \textit{subspaces}, that is, every event corresponds to a subspace of a larger, perhaps infinite dimensional, Hilbert's space. As in \ac{CP}, a subspace is nothing other than an abstraction of an event, and the theory of Hilbert's spaces together with its operations are an abstraction of the different ways the events can occur in the real world.  According to that framework, documents, terms, queries, relevance, aboutness are events modeled as subspaces. While sets and \ac{CP} are widely used in IR, subspaces are not employed  at all for modeling events. An exception is~\cite{vanRijsbergen04} where Hilbert's spaces were introduced for establishing a new theoretical framework encompassing different models proposed by the IR community over the years.
 
What is important to note is that the operations commonly defined for sets are not always defined for subspaces; for example, the notion of union thought for sets cannot be defined on subspaces because the union of two subspaces is not a subspace, while the linear span of two subspaces is. The opposite holds too, i.e., the linear span which is defined for subspaces cannot be defined for sets.  However, there may be analogies~---~the complement of a set corresponds to an orthogonal subspace and both may model the negation of an event, yet the complement of a subspace is not the same as the complement of a set: the former is the set of vectors orthogonal to the vectors of the set, the latter is the set of vectors not in the set. 

When probability is entered, the subspaces used for modeling the events of a real world or an experiment are subjected to a measure, that is, a real function of the subspaces of a Hilbert's space.  The measure is then used for computing the probability of the events modeled by the subspaces.  In this way, the uncertainty of the occurrence of an event can be measured even when sets are replaced with subspaces. In the \ac{QP} case, therefore, the probability space is given by subspaces and a ``\ac{QP}'' function, as illustrated in the following.

As before, let us consider a mathematical formulation. According \ac{QP}, the events $X, Y, \ldots$ are modeled as subspaces $\ket{X}, \ket{Y}, \ldots$ of the space.\footnote{What follows concentrates on one-dimensional subspaces, that is, on the set of vectors spanned by a vector $\ket{X}$ which models an event $A$~---~from now on, $\ket{Y}$ means the subspace too. $\ket{.}$ is called bra-ket notation and was introduced by P. Dirac in {\em The Principles of Quantum Mechanics}, Oxford University Press, 1958.} The definition of the probability of an event modeled by a subspace is different from that of the same event but modeled by a subset. Indeed, the probability of an event modeled by a subspace depends on a subspace and can then be considered as a conditional probability defined as
\begin{eqnarray}
  \Pr(X  |  Y) = |\braket{Y}{X}|^2\ .
  \label{eq:h-prob}
\end{eqnarray}
where the modulus of the inner product between the two vectors is called \textit{amplitude}.  In IR terms, Equation~\ref{eq:h-prob} is the probability that, say, a document described by index term $Y$ is relevant.  The same applies when $X,Y,Z,...$ models, for example,
terms, aboutness, or document clusters.

The conditional probabilities can be defined for both \ac{CP} and \ac{QP}, yet they are defined in different ways. What is different is that \acp{QP} are inherently conditional because the probability of an event is conditioned to a subspace, which refers to another event. Moreover, the conditional probabilities in \ac{QP} theory are symmetric since
\begin{eqnarray*}
  \Pr(X  |  Y) = |\braket{Y}{X}|^2 = |\braket{X}{Y}|^2 = \Pr(Y  |  X)\ .
\end{eqnarray*}
Therefore, the conditional probabilities used in the article become
\begin{equation}
  \label{eq:hilbertian-probabilities}
  p = |\braket{A}{B}|^2 \qquad q = |\braket{B}{C}|^2 \qquad r = |\braket{A}{C}|^2 
\end{equation}

\section{Inequalities of Probability}
\label{app:inequalities}

Suppose there are three observables with $n$ possible values each. Without loss of generality, and for making illustration easier, it is assumed that $n=2$, that is, each observable has two mutually exclusive values, e.g., $\b{A},A$. As for IR, $B$ might be the event that a term is observed in a document, while $\b{B}$ is the event that the term is not observed and $A$ ($\b A$) may denote the event that a document is (not) relevant and so on. Of course, three observables only is a small example, but the fact that the single event space cannot be admitted even when only three observables are examined suggests that it cannot be admitted in more general cases.
%Recall that the conditional \ac{QP} is inherently
%symmetric. Since our interest is the comparison between K-%probability and \ac{QP}, 
It is assumed that
\begin{eqnarray*}
  p = \Pr(A  |  B) = \Pr(B  |  A)\\
  q = \Pr(B  |  C) = \Pr(C  |  B)\\
  r = \Pr(A  |  C) = \Pr(C  |  A)
\end{eqnarray*}
%also when the \ac{CP} is used for computing $p,q,r$,
as in~\cite{Accardi&82}. When \ac{CP} is used, the symmetry of the conditional probability implies that
\begin{eqnarray*}
  \mu(A) = \mu(B) = \mu(C)\ .
\end{eqnarray*} 
This ``symmetry'' may well happen in an IR context; for example, the probability that a given index term is chosed for a relevant document, that is, $\Pr(C | A)$, may equal the probability that the document is assessed as relevant if it has been indexed by the index term.  The fact that the conditional probabilities are usually asymmetric is due to the use of \ac{CP} and then of \ac{BP} which in fact makes $\Pr(A | B)$ different from $\Pr(B | A)$. When \ac{CP} is used, one measure $\mu$ exists such that
\begin{equation}
  p = \frac{\mu(A \cap B)}{\mu(B)} \qquad  q = \frac{\mu(B \cap C)}{\mu(C)}
  \qquad  p = \frac{\mu(A \cap C)}{\mu(C)}\ .
  \label{eq:example-0}
\end{equation}
Suppose, for example, that $p = q = r = \frac{1}{2}$ and $\mu(A) = \mu(B) = \mu(C) = \frac{1}{2}$.  It can easyly be seen that the measures of the co-occurring events can be
computed as
\begin{eqnarray*}
  \mu(A \cap B) = \mu(B \cap C) = \mu(C \cap A) = \frac{1}{4}\ .
\end{eqnarray*}
Things change when other values of $p,q,r$ are estimated from sources independent of each other, for example, as
\begin{eqnarray}
  p = \frac{13}{18} \qquad q = \frac{5}{18} \qquad r = \frac{10}{17}
  \label{eq:example-2}
\end{eqnarray}
The surprising result is that when the values of Equations~\ref{eq:example-2} are considered, a measure $\mu$ cannot be defined in a way such that the probability of $A \cap B$, $B \cap C$, $C \cap A$ and $A \cap B \cap C$ exist~---~no single event space can admit those values of $p,q,r$. The values of Equations~\ref{eq:example-2} are not the only possible values and an infinite number of values of $p,q,r$ exist such that the events do not admit a meausure according to \ac{CP}.  
%What
%happened is that either the observable $A, B$ or $C$ has been observed
%in the condition that a value of another observable has been observed.
%The probability of this conditioned event has then been estimated.
%The same estimation has been for every observable thus obtaining three
%conditioned probabilities $p,q,r$. For infinite values of $p,q,r$, it
%is impossible that these probabilities can be admitted in the
%K-theory of probability. 
As \ac{CP} estimates the conditional probabilities on the basis of \ac{BP}, it follows that,
\begin{equation*}
  \Pr(A | B) \neq \frac{\mu(A \cap B)}{\mu(B)}
\end{equation*}
since $\mu(A \cap B)$ cannot be calculated. If the co-occurrence of
events, e.g. $A \cap B$ was observed, and the frequency of these co-occurring events were available, an estimation of $\mu(A \cap B)$ would be possible thus making \ac{BP} valid.  When $\mu$ cannot be defined for some observed conditional probabilities, one has to conclude that the co-occurrence of events is impossible, namely, statements like ``both $A$ and $B$ occur'' do not make any sense.  The inequality which acts as the test of the existence of a single event is proven in~\cite{Accardi&82} and is stated as \begin{proposition}
  $p,q,r$ admit a single event space if and only if 
  \begin{equation}
    |p + q - 1| \leq r \leq 1 - |p - q|
    \label{eq:accardi-1}
  \end{equation}
  \label{prop:accardi-1}
\end{proposition}
When Inequality~\ref{eq:accardi-1} is violated, neither measure $\mu$ nor sets $A,B,C, ...$ can be defined for the observables $A,B,C$ such that \ac{BP} holds.

Inequality~\ref{eq:accardi-1} provides for a simple test to check if the conditional probabilities estimated in different experimental conditions are compatible with a single event space. The question is then, what is the probability space if that single event is incompatible? The answer was provided in~\cite{Accardi&82} and is reported here without proof.
\begin{proposition}
  $p,q,r$ admit a complex \ac{QP} space if and only if 
  \begin{equation}
    \left(\sqrt{pq} - \sqrt{1-p}\sqrt{1-q}\right)^2 \leq r \leq \left(\sqrt{pq} + \sqrt{1-p}\sqrt{1-q}\right)^2
    \label{eq:accardi-2}
  \end{equation}
  \label{prop:accardi-2}
\end{proposition}
\begin{proposition}
  $p,q,r$ admit a real \ac{QP} space if and only if 
  \begin{equation}
    r=\left(\sqrt{pq} - \sqrt{1-p}\sqrt{1-q}\right)^2 \qquad \mbox{or}
    \qquad r=\left(\sqrt{pq} + \sqrt{1-p}\sqrt{1-q}\right)^2 
    \label{eq:accardi-3}
  \end{equation}
  \label{prop:accardi-3}
\end{proposition}
From these inequalities, some simple results follow. First, if $p,q,r$ admit a single event space, then they also admit a \ac{QS}. Second, there are infinite values of $p,q,r$ which admit a complex \ac{QS}, and not a single event space.  When $p,q,r$ admit a complex (real) \ac{QP} space, the events $A,B,C$ can be represented as vectors $\ket{A}, \ket{B}, \ket{C}$ of a complex (real) \ac{QS} such that the probabilities are those defined in Section~\ref{app:hilbert}.  There are cases when either the complex \ac{QS} nor the \ac{CP} space can be admitted; for example, $p=1/10, q=2/10, r=3/10$. This implies that that \ac{QS} is a necessary yet not sufficient framework. 

Up to now, the relatively simple case of three observables or properties has been considered. A more general result states a necessary and sufficient condition that a set of probabilities admit a single event space. This result is due to Pitowsky, was proven in~\cite{Pitowsky89} and is illustrated in Appendix~\ref{app:pitowsky}.

\section{Pitowsky's Theorem}
\label{app:pitowsky}

Suppose $n \geq 2$ properties are observed from a, say, collection of documents~---~in particular, the case $n=3$ was considered in the previous sections where the properties were labeled as $A,B,C$ and their respective negations $\b A,\b B,\b C$ to mean, for example, that a document was relevant ($A$) or not ($\b A$). 

Suppose also that a series of experiments yielded the $n(n+1)/2$ probabilities \begin{equation*}
p(1), \ldots, p(n), p({1,1}), \ldots, p({i,j}), \ldots, p({n-1,n})
\end{equation*}
where $1 \leq i < j \leq n$, where $p_i$ is the probability that the event $A_i$ occurs and $p_{i,j}$ is the probability that events $A_i, A_j$ are observed in the series of experiments.  These probabilities can be arranged in the correlation vector $\textbf{p}$. Given $\textbf{p}$, under what conditions a single single event space for the events $A_1, \ldots, A_n$ and the measure $\mu$ can be defined such that for all $i,j$ where $1 \leq i < j \leq n$
\begin{equation*}
  p(i) = \mu(A_i) \qquad p({i,j}) = \mu(A_i \cap A_j)\ ?
\end{equation*}
The answer was provided in~\cite{Pitowsky89}.

Suppose that $n$ properties are considered and that all the strings of $n$ binary numbers $0$'s and $1$'s are built. Let $b$ be one of these binary strings, i.e. $b \in \{0,1\}^n$; for example, when $n=2$, then $b = \texttt{01}$ is such a string. The binary digit $\texttt{1}$ means that $A_i$ was observed, while $\texttt{0}$ means that it was not.

Once $b$ is fixed, the correlation vector of $n(n+1)/2$ probabilities $\mathbf{p}_b$ is defined as follows: $\mathbf{p}_b(i) = b_i$ and $\mathbf{p}_b({i,j}) = b_ib_j$; for example, when $b = \texttt{01}$, then $\mathbf{p}_b = (0,1,0)$.  There are $2^n$ such correlation vectors since $2^n$ binary strings can be enumerated using $n$ binary digits.

The theorem defined the closed, convex polytope whose vertices are the $2^n$ correlation vectors like $\mathbf{p}_b$, that is, the polytope is the set of all points that can be expressed by a linear combination of these $2^n$ correlation vectors. As formula, the polytope is expressed as \begin{equation*}
  \textbf{p} = \lambda_1\mathbf{p}_{b_1} + \cdots \lambda_{2^n}\mathbf{p}_{b_{2^n}}
\end{equation*}
where $b_i$ is the $i$-th binary string, $\mathbf{p}_{b_i}$ is the
correlation vector of probabilities built from $b_i$ and \begin{equation*}
  \lambda_i \geq 0 \qquad \sum_{i=1}^{2^n} \lambda_i = 1\ .
\end{equation*}
Then, the following proposition holds
\begin{theorem}
  For all $n$ and all correlation vectors of probabilities $\textbf{p}$,   $\textbf{p}$ admits a single event space of $n$ (not necessarily   distinct) events if and only if $\textbf{p}$ belongs to the   polytope.
  \label{prop:pitowsky-0}
\end{theorem}
This means that the $2^n$-unknowns system of $n(n+1)/2$ linear equations
\begin{equation}
  \left\{
    \begin{array}{cclcccl}
      p(1)    & = & \lambda_1p_{b_1}(1) & + & \cdots & + & \lambda_{2^n}p_{b_{2^n}}(1)\\
      \vdots  & = & & & \vdots \\
      p(n)    & = & \lambda_1p_{b_1}(n) & + & \cdots & + & \lambda_{2^n}p_{b_{2^n}}(n)\\
      p({1,n})& = & \lambda_1p_{b_1}({1,n}) & + & \cdots & + & \lambda_{2^n}p_{b_{2^n}}({1,n})\\
      \vdots  & = & & & \vdots \\
      p({m,n})& = & \lambda_1p_{b_1}({m,n}) & + & \cdots & + & \lambda_{2^n}p_{b_{2^n}}({m,n})\\
    \end{array}
  \right.
  \label{eq:system}
\end{equation}
have solutions if and only if $\textbf{p}$ belongs to the polytope.

Note that since the events are not necessarily distinct, one can assume that when $n=2$, $A_1 = A_2$ and therefore $p(1)=p(2)=p(1,2)$. In this way, the theorem always holds since $\mu(A_1) = p(1)$ and the event space includes only one observable.

The theorem holds if and only if a system of inequalities holds. For example, suppose $n=2$, $n(n+1)/2=3$ and $2^2=4$. The binary strings are $b_1=\texttt{000}, b_2=\texttt{010}, b_3=\texttt{100}, b_4=\texttt{111}$. Three observed probabilities $p(1), p(2), p(1,2)$ admit a single event space if and only if the system:
\begin{equation*}
  \left\{
    \begin{array}{l}
      0 \leq p(1,2) \leq p(1) \leq 1 \\
      0 \leq p(1,2) \leq p(2) \leq 1 \\
      0 \leq p(1)+p(2)-p(1,2) \leq 1 \\
    \end{array}
  \right.
\end{equation*}
admits solutions. For example, no solutions can be admitted when $p({1,2}) > p(1)$ or $p({1,2}) > p(2)$. When $n=3$,
\begin{equation*}
  \left\{
    \begin{array}{ll}
      0 \leq p(i,j) \leq p(i) \leq 1 & 1 \leq i < j \leq 3 \\
      0 \leq p(i)+p(j)-p(i,j) \leq 1 & 1 \leq i < j \leq 3 \\
      p(1) - p(1,2) - p(1,3) + p(2,3) \geq 0 \\
      p(2) - p(1,2) - p(2,3) + p(1,3) \geq 0 \\
      p(3) - p(1,3) - p(2,3) + p(1,2) \geq 0 \\
    \end{array}
  \right.
\end{equation*}

The main problem is that a polynomial-time algorithm which tests if
$\textbf{p}$ belongs to the polytope for every $n$ does not exist. However, in this article, our interest is not to compute the polytope, but to have theoretical results which provide the necessary and sufficient conditions so that $\textbf{p}$ admits a single event space.

\end{document}